\providecommand{\arxiv}[1]{#1}
\providecommand{\acm}[1]{}

\arxiv{
\documentclass[12pt]{article}
\usepackage{fullpage}
\newcommand\keywords[1]{\noindent{\bf keywords:} #1}
\usepackage{xcolor}
}

\acm{
\documentclass[sigconf]{acmart}
\AtBeginDocument{%
  }

\copyrightyear{2023}
\acmYear{2023}
\setcopyright{acmlicensed}\acmConference[ApPLIED 2023]{The 5th workshop on Advanced tools, programming languages, and PLatforms for Implementing and Evaluating algorithms for Distributed systems}{June 19, 2023}{Orlando, FL, USA}
\acmBooktitle{The 5th workshop on Advanced tools, programming languages, and PLatforms for Implementing and Evaluating algorithms for Distributed systems (ApPLIED 2023), June 19, 2023, Orlando, FL, USA}
\acmPrice{15.00}
\acmDOI{10.1145/3584684.3597275}
\acmISBN{979-8-4007-0128-3/23/06}
}

\usepackage{graphicx} 
\usepackage{appendix}
\usepackage{listings}
\usepackage{alltt}
\arxiv{
    \usepackage{url}
}
\newcommand{\Hex}[1]{\hspace{#1ex}}
\newcommand{\Vex}[1]{\vspace{#1ex}}
\newcommand{\mypar}[1]{\Vex{2}\noindent{\bf #1.~}\Vex{.0}}
\newcommand{\mysubpar}[1]{\Vex{.0}\Hex{-0}{\bf #1.}\Vex{-.0}}
\usepackage{enumitem}
\setlist{leftmargin=\arxiv{6}\acm{4}ex, itemsep=.6ex}
\newenvironment{code}{\Vex{-.0}\begin{alltt}\small}{\end{alltt}\Vex{-.0}}
\newcommand\co[1]{\mbox{\small\tt #1}} 
\newcommand\coline[1]{\Vex{1}\\\co{\Hex{3}#1}\Vex{1}\\} %
\newcommand\m[1]{$#1$} %
\newcommand\p[1]{${\it #1}$} %
\newcommand{\eg}[1]{}

\definecolor{codegreen}{rgb}{0,0.6,0}
\definecolor{codegray}{rgb}{0.5,0.5,0.5}
\definecolor{codepurple}{rgb}{0.58,0,0.82}
\definecolor{backcolour}{rgb}{0.95,0.95,0.92}

\makeatletter
\lst@Key{countblanklines}{true}[t]%
    {\lstKV@SetIf{#1}\lst@ifcountblanklines}

\lst@AddToHook{OnEmptyLine}{%
    \lst@ifnumberblanklines\else%
       \lst@ifcountblanklines\else%
         \advance\c@lstnumber-\@ne\relax%
       \fi%
    \fi}
\makeatother

\lstdefinestyle{basestyle}{
    basicstyle=\ttfamily\scriptsize,
    breakatwhitespace=false,         
    breaklines=true,                 
    captionpos=b,                    
    keepspaces=true,
    numbers=left,
    numbersep=5pt,                  
    showspaces=false,                
    showstringspaces=false,
    showtabs=false,
    numberblanklines=false
}

\lstdefinestyle{pythonstyle}{
    style=basestyle, 
    commentstyle=\color{codegreen},
    keywordstyle=\color{magenta},
    numberstyle=\tiny\color{codegray},
    stringstyle=\color{codepurple},
    language=python,
    countblanklines=false
}

\lstdefinestyle{pseudocodestyle}{
    style = basestyle,
    basicstyle=\ttfamily\footnotesize,
    commentstyle=\color{black},
    keywordstyle=\color{blue},
    numberstyle=\tiny\color{black},
    stringstyle=\color{black}
}

\acm{
\lstdefinestyle{outputstyle}{
    style = basestyle,
    commentstyle=\color{black},
    keywordstyle=\color{magenta},
    numberstyle=\tiny\color{black},
    stringstyle=\color{black}
}
}

\arxiv{
\lstdefinestyle{outputstyle}{
    style = basestyle,
     basicstyle=\ttfamily\tiny,
    commentstyle=\color{black},
    keywordstyle=\color{magenta},
    numberstyle=\tiny\color{black},
    stringstyle=\color{black}
}
}

\acm{
    \lstset{style=pythonstyle}
}

\arxiv{
    \lstset{style=pythonstyle, basicstyle=\ttfamily\tiny}
}
\arxiv{
\title{
Specification and Runtime Checking of Derecho,
A Protocol for Fast Replication for Cloud Services}
}

\acm{
\title[Specification and Runtime Checking of Derecho]{
Specification and Runtime Checking of Derecho,
A Protocol for Fast Replication for Cloud Services}
}

\arxiv{
\author{
Kumar Shivam  \Hex{5} 
Vishnu Paladugu \Hex{5}
Yanhong A. Liu \\
Stony Brook University\\
\{kshivam,vpaladugu,liu\}@cs.stonybrook.edu
}
\date{May 17, 2023}
}

\acm{
\author{Kumar Shivam}
\email{kshivam@cs.stonybrook.edu}
\affiliation{%
  \institution{Stony Brook University}
  \country{USA}
}

\author{Vishnu Paladugu}
\email{vpaladugu@cs.stonybrook.edu}
\affiliation{%
  \institution{Stony Brook University}
  \country{USA}
}

\author{Yanhong A. Liu}
\email{liu@cs.stonybrook.edu}
\affiliation{%
  \institution{Stony Brook University}
  \country{USA}
}
}

\begin{document}
\arxiv{\maketitle}

\begin{abstract}
Reliable distributed systems require replication and
consensus among distributed processes to tolerate process and
communication failures. Understanding and assuring the correctness of
protocols for replication and consensus have been a significant
challenge.
This paper describes the precise specification and runtime checking of
Derecho, a more recent, sophisticated protocol for fast replication and consensus for cloud services.

A precise specification must fill in missing details and resolve
ambiguities in English and pseudocode algorithm descriptions
while also faithfully following the descriptions.
To help check the correctness of the protocol,
we also performed careful manual analysis and
increasingly systematic runtime checking.
We obtain a complete specification that is directly executable, 
and we discover and fix a number of 
issues in the pseudocode. These results were facilitated by the already detailed pseudocode of Derecho
and made possible by using DistAlgo, a
language that allows distributed algorithms to be easily and clearly expressed
and directly executed.
\end{abstract}

\acm{

}


\acm{\maketitle}

\section{Introduction}

Reliable distributed systems require replication and consensus
among distributed processes to tolerate process and communication
failures.
Many algorithms and variations have been proposed for replication and consensus, starting from Virtual Synchrony (VS) by Birman and Joseph \cite{birman1987exploiting}, and Viewstamped Replication (VR) by Oki and Liskov \cite{VR}, including the well-known Paxos algorithm by Lamport~\cite{partTimeParliament}, among many others, e.g., see~\cite{vra-2015,ChaLiu20Liveness-PODC21}.
However, understanding and assuring the correctness of these algorithms have remained a significant challenge, especially as more sophisticated algorithms are being developed.

\mypar{This paper} This paper describes the precise specification and runtime checking
of Derecho~\cite{Jha2019DerechoFS}, a more recent, sophisticated protocol for fast replication
and consensus for cloud services.
%
Derecho provides state machine replication and dynamic membership tracking,
especially
for replicating large data with non-blocking pipelines, and is shown to be significantly faster than comparable widely used,
highly-tuned, standard tools.
It employs a lock-free distributed shared
memory called a shared-state table (SST) for sharing protocol-control information, especially suitable for running on remote direct memory access (RDMA).

Our specification is written in DistAlgo~\cite{Liu+17DistPL-TOPLAS}, a language that allows distributed algorithms to
be easily and clearly expressed and directly executed. 
It provides all three benefits enabled by DistAlgo: (1) distributed processes and communications, both synchronous and asynchronous, are expressed at a high level as pseudocode, (2) the specification is completely precise, supported by formal operational semantics of DistAlgo, and (3) the specification is directly executable in distributed environments, supported by the DistAlgo compiler that is built on top of the Python compiler.

A precise specification must fill in missing details and resolve ambiguities in English and pseudocode algorithm descriptions while
also faithfully following the descriptions. 
Our specification is especially facilitated by Derecho's already detailed pseudocode and descriptions~\cite[Appendix A]{Jha2019DerechoFS}, as well as Derecho's active team of experienced researchers and developers. Derecho pseudocode uses exact fields of structures for information kept in the system state, especially including for the SST, and provides in detail the key steps in both steady-state execution and view change protocols.

To help check the correctness of the Derecho specification, we also performed careful manual analysis and
increasingly systematic runtime checking. 
The clarity of the specification allows some issues to be noticed by quick manual inspection, while automatic running and checking allow more subtle issues to be discovered.

We specified and checked well-established safety properties as well as 
various progress queries and results.
These specifications and automatic checking are enabled by a general framework for runtime checking of safety and liveness properties~\cite{RV2020} supported by DistAlgo.  As a result, the properties are specified at a high level as logical statements and are checked automatically by a checker process while the protocol runs, without changes to the specification of the protocol that can obscure the clarity of the protocol specification.

There has been a significant amount of related research, as discussed in Section~\ref{sec-related}. Our work contains three main contributions:
\begin{itemize}
\item We develop a rigorous specification of Derecho that corresponds closely to the pseudocode and is complete, precise, and directly executable.
\item We discover and fix a number of issues in the Derecho pseudocode, e.g.,~\cite[Errata, page 50]{Jha2019DerechoFS-Cornell}, and helped improve the pseudocode~\cite[page 72]{Jha22thesis}. 
\item We demonstrate through Derecho a practical method for developing a rigorous and improved specification through not only manual inspection but also automated runtime checking.
\end{itemize}
Note that the bugs and fixes we found are for the Derecho pseudocode~\cite{Jha2019DerechoFS},
and have been checked and confirmed by the Derecho team~\cite{Jha2019DerechoFS-Cornell,Jha22thesis,Jha23email}.
In all cases, Derecho developers have also checked and confirmed that the bugs are not in their implementation in C++~\cite{Jha22thesis,Jha23email}.

Bugs in protocol pseudocode are quite normal, simply because
pseudocode is manually created and there is no way to run or
check other than by staring at it.
DistAlgo is exactly for expressing protocols easily and precisely at such pseudocode level, and then running them for testing and for systematic runtime checking of desired properties.
%
A complete specification of Derecho in DistAlgo can be found at~\cite{derecho-da-github}.

\section{Derecho and specification language}

\subsection{Derecho overview}



Derecho \cite{Jha2019DerechoFS} is a replication protocol for coordinating distributed actions.
The protocol supports state machine replication by utilizing
specialized hardware technology, specifically RDMA. RDMA enables direct access to remote memory without involving the CPU, using hardware such as Network Interface Card (NIC), resulting in higher throughput and lower latency through avoidance of context switching.


The protocol employs group multicasting to order client request messages and supports atomic multicast and total-ordered message delivery. 
In this context, a group is defined as a set of member processes referred to as nodes.
Atomic multicast ensures that messages sent by a member node are either delivered to all member nodes or none at all, while total-ordered message delivery guarantees that messages are delivered in the same order to all member nodes in the group.


Each node in a group maintains a copy of SST, one row for each node.
Each node updates its own row in the SST and propagates the update to other nodes using RDMA.

The protocol has two main parts.

\mypar{(1) 
Steady-state execution} 
Derecho employs the SST multicast (SMC) protocol for small message multicast, as described in Section \ref{sec-steady}. To initiate a multicast, one of the nodes in the group stores the incoming request message in SST's ring buffer data structure, 
which is propagated to all nodes in the group.
Each node buffers the message upon receiving it until it knows all the nodes in the group have received it.
Atomic multicast delivery of a message happens when all the previous messages have been delivered on all nodes, and the current message has been received on all nodes.


\mypar{(2) View change}
Derecho employs virtual synchrony \cite{birman1987exploiting} to track dynamic membership in a process group computing style. 
Process groups allow members to join and leave the group while the application is active, triggering a membership change.
Each membership change initiates a new epoch (view), and the protocol progresses through a series of epochs, each with its own membership.
A two-phase commit pattern is used to carry out a membership change, with information exchanged via the SST.


\subsection{Language for precise specification}

For precise executable specification of Derecho at a high level that corresponds to algorithm pseudocode,
we use DistAlgo~\cite{Liu+12DistPL-OOPSLA,Liu+17DistPL-TOPLAS}.
%
DistAlgo supports the following four main concepts of distributed programming by building on an object-oriented programming language,
Python.

\mypar{(1) Distributed processes that can send messages}
A type \co{\p{P}} of processes is defined by
\begin{code}
    process \p{P}: \p{stmt}
\end{code}
The body \p{stmt} may contain, among usual definitions,
\begin{itemize}
\setlength{\itemsep}{0ex}

\item a \co{setup} definition for \arxiv{}setting up the
  values used in the process, %

\item a \co{run} definition for running the main %
  flow of the process\eg{}, and

\item \co{receive} definitions for handling received messages\eg{}.

\end{itemize}
A process can refer to itself as \co{self}. Expression \co{self.\p{attr}}
(or \co{\p{attr}} when there is no ambiguity) refers to the value of
\co{\p{attr}} in the process.
\begin{itemize}
\setlength{\itemsep}{0ex}

\item \co{\p{ps} := \p{n} new \p{P}}
  creates \co{\p{n}} new processes of type \co{\p{P}},
  and assigns the new processes to
  \co{\p{ps}}\eg{}.

\item %
\co{\p{ps}.setup(\p{args})} sets up processes \co{\p{ps}} using values of
\co{\p{args}}\eg{}.

\item \co{\p{ps}.start()} starts \co{run} of \co{\p{ps}}\eg{}.

\end{itemize}
\co{new} can have an additional clause, \co{at \p{node}},
specifying remote nodes where the created processes will run; the default
is the local node.

A process can easily send %
a message \co{\p{m}} to processes \co{\p{ps}}:
\begin{code}
    send \p{m} to \p{ps}
\end{code}\Vex{-1.2}

\mypar{(2) Control flow for handling received messages}
Received messages can be handled both asynchronously, using 
\co{receive} definitions, and synchronously, using \co{await}
statements.
\begin{itemize}

\item A \co{receive} definition is of the following form: \coline{receive
    \p{m} from \p{p}:~\p{stmt}} %
  It handles, at yield points, un-handled messages that match \co{\p{m}}
  from \co{\p{p}}\eg{}.
  A yield point is of the form \co{-\,-\,\p{l}}, where \p{l} is a label; it specifies a point in the program where control yields to handling of un-handled messages, if any, and resumes afterwards.
  There is an implicit yield point before each \co{await} statement\eg{}, for handling messages while waiting.
  The \co{from} clause is optional.

\item An \co{await} statement is of the following form: \coline{await
    \p{cond_1}:\,\p{stmt_1} or\,...\,or \p{cond_k}:\,\p{stmt_k}
    timeout\,\p{t}:\,\p{stmt}} %
  It waits for one of \co{\p{cond_1}}, ..., \co{\p{cond_k}} to be true or a
  timeout after period \co{\p{t}}, and then nondeterministically selects
  one of \co{\p{stmt_1}}, ..., \co{\p{stmt_k}}, \co{\p{stmt}} whose
  conditions are true to execute\eg{}.
  Each branch is optional.
  So is the statement in \co{await} with a single branch.
\end{itemize}\Vex{-1.2}

\mypar{(3) High-level queries for synchronization conditions}
High-level queries can be used over message histories, and patterns can be
used to match messages.
\begin{itemize}
  \setlength{\parskip}{.5ex}

\item Histories of messages sent and received by a process are kept in
  \co{sent} and \co{received}, respectively.
  \co{sent} is updated at each \co{send} statement, by adding each message
  sent.
  \co{received} is updated at the next yield point if there are un-handled
  messages, by adding un-handled messages before executing all matching
  \co{receive} definitions.

  Expression \co{sent \m{m} to \p{p}} is equivalent to \co{\p{m} to \p{p} in
    sent}.  It returns true iff a message that matches \co{\p{m} to \p{p}}
  is in \co{sent}.
  The \co{to} clause is optional.
  Expression \arxiv{\linebreak}\co{received \m{m} from \p{p}} is similar.

\item A pattern can be used to match a message, in \co{sent} and
  \co{received}, and by a \co{receive} definition.  A constant value, such
  as \co{"release"}, or a previously bound variable, indicated with prefix
  \co{=}, in the pattern must match the corresponding components of the
  message.  An underscore \co{\_} matches anything.  Previously unbound
  variables in the pattern are bound to the corresponding components in the
  matched message.

  For example, \co{received("release",t3,=p2)} %
  matches every triple in \co{received} whose first component is
  \co{"release"} and third component is the value of \co{p2}, and
  binds \co{t3} to the second component.

\end{itemize}
A query can be an existential or universal quantification, a comprehension,
or an aggregation over sets or sequences.
\begin{itemize}

\item An existential quantification and a universal quantification are of
  the following two forms, respectively: \coline{some \p{v\sb{1}} in
    \p{s\sb{1}}, ..., \p{v\sb{k}} in \p{s\sb{k}} has \p{cond}}\Vex{-5}
  \coline{each \p{v\sb{1}} in \p{s\sb{1}}, ..., \p{v\sb{k}} in \p{s\sb{k}}
    has \p{cond}} %
  They return true iff for some or each, respectively, a combination of
  values of variables that satisfies all \co{\p{v\sb{i}} in \p{s\sb{i}}}
  clauses, \co{\p{cond}} holds\eg{}.

\item A comprehension is of the following form:
  \coline{\{\p{e}:~\p{v\sb{1}} in \p{s\sb{1}}, ..., \p{v\sb{k}} in
    \p{s\sb{k}}, \p{cond}\}} %
  It returns the set of values of \co{\p{e}} for all combinations of values
  of variables that satisfy all \co{\p{v_i} in \p{s\sb{i}}} clauses and
  condition \co{\p{cond}}\eg{}.

\item An aggregation is of the form \co{\p{agg} \p{s}}, where \co{\p{agg}}
  is an aggregation operator such as \co{count} or \co{max}.  It returns
  the value of applying \co{\p{agg}} to the set value of \co{s}\eg{}.

\item In all query forms above, each \co{\p{v_i}} can be a pattern.

\end{itemize}
Other operations, such as set union and sequence concatenation, can also be used.

\mypar{(4) Configuration for setting up and running}
Configuration for requirements such as the use of logical clocks and the use of
reliable and FIFO channels can be specified in a \co{main} definition.
For example, \co{configure channel = fifo} specifies that fifo channels are used and TCP is used for process communication. 

DistAlgo also supports automatic visualization of replays forward and backward, making it much easier to understand protocol runs.

\mypar{DistAlgo compiler, Python syntax, queries, and extended \co{sent} and \co{received}}
To allow anyone with Python to run DistAlgo directly,
DistAlgo compiler supports the Python syntax~\cite{Distalgo17lang}.
For example, \co{send m to p} is written as \co{send(m, to=p)}, and\linebreak
\co{each sent m to p has cond} is written as \co{each(sent(m, to=p),}\acm{\linebreak} \co{has=cond)}; in patterns, \co{=var} is written as \co{\_var}. 

While Derecho pseudocode does not use high-level set queries, 
these queries are critical to specify the many reducer functions used.
Also, slightly extended
\co{p.sent} and \co{p.received}, 
denoting the
process \co{p}'s \co{sent} and \co{received} sequences, respectively, is critically helpful for specifying the properties to be checked.

In our specification in DistAlgo, the following convention for comments are used:
(1) comments after \co{\#} are text or pseudocode copied from the Derecho paper~\cite{Jha2019DerechoFS}, 
except when noted as from email with Sagar Jha;
(2) comments after \co{\#\#} (or no comments) describe code we had to fill in; and
(3) comments after \co{\#\#\#} describe changes to the pseudocode in paper.


\section{Specifying system state} 

Information maintained in a system is essential for specifying the system.
We define classes with fields that allow the algorithm steps in DistAlgo to match the corresponding steps in pseudocode exactly.
Fig.~\ref{fig-system-state} shows the complete precise specification in DistAlgo.

The key information maintained by Derecho is the SST, specified as a list of \co{SSTRow} objects. Fig.~\ref{fig-system-state} (lines 7-32) shows the definition of class \co{SSTRow} with its fields.
For instance,  field \co{slots} (line 13) is a list for a ring buffer, with reusable slots (lines 1-6) for request messages and related metadata.

In addition to the SST, Derecho needs a \co{View} object to hold information about an epoch (used interchangeably with view in Derecho), such as the leader and members in the view. This object is critical in the membership-change protocol, which is triggered when a member joins or leaves the group. 
Fig.~\ref{fig-system-state} shows the definition of class \co{View} (lines 33-47) with its fields and methods to add and remove members. 



\begin{figure*}
\begin{lstlisting}[firstline=65,lastline=115]
class Slot:
  """A slot stores a client request message. A vector of slots is a field in SST"""
  def __init__(self):
    self.buf = None  # (p.33) a request of up to max_message_size characters  ## initialized in Node.init()
    self.index = 0   # (p.33) index associated with a slot  ## initialized in Node.init()
    self.size = 0    # ## size of the request, set in get_buffer()  ### defined but not used

class SSTRow:
  """A shared state table (SST) row that stores info about a node.
     A SST has a SSTRow for each member node and is stored in each member"""
  def __init__(self, n, window_size):  ## initialize SST columns for the row; all used in pseudocode but the last two 
    # n:           ## number of member nodes in the group
    # window_size: ## length of vector of slots for storing client req msgs received by the node, directly or indirectly
    self.slots = [Slot() for _ in range(window_size)]  # (p.33) vector of window_size slots  ## for client request msgs
    self.received_num = [-1] * n                       # (p.33) number of messages received from each node  ### number-1
                                                         ## initialized in Node.init(), and set in receive_req()
    self.global_index = -1                 # ## global index of last message received from the most lagging node
    self.latest_delivered_index = -1       # ## min of self.global_index over all members
    self.latest_received_index = [-1] * n  # ## index of latest msg received from each node, set in recv to received_num
                                             ## i.e., self.received_num-1  ### redundant, but not clear with null msgs
    self.min_latest_received = [-1] * n    # ## for each node, min of latest_received_index over all rows in SST
    self.suspected = [False] * n           # ## for each node, whether that node is suspected to have failed
    self.wedged = False     # ## true when any node is suspected 
    self.changes = []       # ## list of nodes suspected (or added from joins) to proposed as changes by the leader
    self.num_changes = 0    # ## number of nodes in self.changes, i.e., length of self.changes
    self.num_acked = 0      # ## number of nodes in self.changes acknowledged
                              ### set in 1 place by us, using num_changes
    self.num_committed = 0  # ## min of self.num_acked over not suspected nodes
    self.num_installed = 0  # ## number of nodes installed (added/removed) by the node, as proposed by the leader
    self.ragged_edge_computed = False  # ## true for leader calling terminate_epoch or others after leader did;
                                         ## the call happens when leader's num_committed > self's num_installed
    self.active = False  # (p.40) ## true when the epoch is active, only used at start  ### not in pseudocode 
                            ### could use logical or over sst[my_rank].suspected or even just own suspected 

class View(): 
  """A view that holds the information of an epoch. An epoch is the duration of a view."""
  def __init__(self, n, epoch=0, leader_rank=0):
    # n:  ## number of members in the view
    self.epoch = epoch              # ## epoch number of the view; epoch and view used interchangeably in the paper
    self.leader_rank = leader_rank  # ## index of the leader 
    self.members = [None] * n       # ## list of member nodes in the view
    self.failed = [False] * n       # ## for each node, whether that node is suspected and thus considered failed

  def add_member(self, node):         ## add member to the view
    self.members.append(node)         ## append node to members
    self.failed.append(False)         ## add the failed attribute corresponding to the added node

  def remove_member(self, node):      ## remove node from members of the view
    index = self.members.index(node)  ## get index of node in members
    del self.failed[index]            ## remove failed entry for node
    self.members.remove(node)         ## remove node from members
\end{lstlisting}
\Vex{-2}
\caption{Specification of system state.}
\label{fig-system-state}
\end{figure*}

Function \co{write\_sst} in Fig.~\ref{fig-write-sst} (lines 1-5) specifies an update to the SST. 
It updates the local SST and sends an \co{rdma\_write\_sst} message to other nodes.
Upon receiving the message, nodes execute a \co{receive} handler (lines 6-17), which
calls function \co{wt\_local\_sst} to update the SST row corresponding to the sender node (lines 18-24).

\begin{figure*}
\begin{lstlisting}[firstline=122,lastline=147]
  def write_sst(row, col, val, index=None):  ## write SST entry at row and col, at index if col is a list, with val
    wt_local_sst(row, col, val, index)                               ## write local SST
    msg = ('rdma_write_sst', row, col, val, index, curr_view.epoch)  ## msg to send to others  
    msg_tagged = ('data' if col == "slots" else 'control',) + msg    ## tag the msg as data or control
    send(msg_tagged, to= others)                                     ## send the message to other nodes
    
  def receive(msg = (_, 'rdma_write_sst', row, col, val, index, epoch)):  ## _ ignores tag 'data' or 'control'
    output("received: ", ('rdma_write_sst', row, col, val, index, epoch))
    if epoch != curr_view.epoch:                  # (p.40) ## if msg is for a different epoch, ignore
      output('received rdma_write_sst msg for different epoch, current: ', curr_view.epoch, ' and msg: ', epoch)
      return
    if col == "slots" and any(curr_view.failed):  # (p.12) ## if msg is a req and curr view has failed members, ignore
      output("Failure detected, new data messages dropped")
      return
    if row in freeze:                             # (p.16) ## if msg is for a row in freeze, ignore; should do at write 
      output("ignored rdma_write_sst msg because of frozen row: ", row, " node: ", G[row])
      return
    wt_local_sst(row, col, val, index)            ## write local SST

  def wt_local_sst(row, col, val, index=None):  ## write local entry at row and col, at index if col is a list, with val
    if index is None:                           ## if index is None, meaning col is not a list
      setattr(sst[row], col, val)               ## just update SST entry with val
    else:                                       ## col is a list
      entry = getattr(sst[row], col)            ## retrieve SST entry at row and col
      entry[index] = val                        ## update the element at index of entry with val
      setattr(sst[row], col, entry)             ## update SST entry at row and col with updated entry
\end{lstlisting}
\Vex{-2}
\caption{Specification of function for writing to SST.}
\label{fig-write-sst}
\end{figure*}

The protocol uses a number of reducer functions on the SST. These functions read SST entries and compute aggregation information. 
These reducer functions are specified to compute exactly as stated in the pseudocode.
Fig.~\ref{fig-reducer-function} shows an example, \co{LogicalOr}.

\begin{figure}
\begin{lstlisting}[firstline=170,lastline=172]
  def LogicalOr(col):
    """logical 'OR' of all values in column col in sst"""
    return any(getattr(sst[row], col) for row in range(len(sst)))
\end{lstlisting}
\Vex{-2}
\caption{Specification of an example reducer function.}
\label{fig-reducer-function}
\end{figure}




\section{Specifying steady-state execution}
\label{sec-steady}
Steady-state execution describes 
atomic multicast delivery
of a client request message across the nodes in a group.
The messages are delivered in a round-robin fashion according to a global message order captured by \co{global\_index}, where each node sends 
one multicast in each delivery cycle.

Fig.~\ref{fig-steady} shows a complete precise specification of steady-state execution in DistAlgo.
It consists of three main parts: sending messages, receiving messages, and in-order delivery of messages.


\begin{figure*}
\Vex{-2}
\begin{lstlisting}[firstline=273,lastline=356]
  def receive(msg=('request', req)):  ## receive request from client, and send req to the group
    client, req_id, _ = req           ## request is of form (client, req_id, cmd)
    output("request: ", req, " received from client: ", client)
    if some(sent(('response', _req_id, res), to= _client)):  ## if a response to req was already sent
      output("Duplicate request received: ", req, ". Sending result: ", res, " back.")
      send(('response', req_id, res), to= client)            ## send the response again
      return
    send_req(req)                     ## send request by putting it in the next slot, if a slot is available

  # (p.33) A.2.3 Sending.  First the sending node reserves one of the slots:
  def get_buffer(msg_size):                       # char* get_buffer(msg_size) {
    assert msg_size <= max_msg_size               #   assert(msg_size <= max_msg_size);
    completed_num = Min("received_num", my_rank)  #   completed_num = Min{sst[*].received_num[my_rank]};
    if sent_num - completed_num == window_size:   #   if (sent_num - completed_num < window_size) { 
      ### changed < to ==  ### first changed < to > in ERRATA of paper, but it caused deadlock
      output("slot verctor full, returning, sent_num: ", sent_num, " completed_num: ", completed_num)
      return None                                 #     return nullptr; }
    slot = (sent_num + 1) % window_size           #   slot = (sent_num + 1) % window_size;
    sst[my_rank].slots[slot].size = msg_size      #   sst[my_rank].slots[slot].size = msg_size;
    return slot                                   #   return sst[my_rank].slots[slot].buf; }  ### return slot
  
  # (p.34) After get_buffer returns a non-null buffer,
  # the application writes the message contents in the buffer and calls send  
  def send_req(req):                              # void send() {
    slot = get_buffer(len(req[2]) if req else 0)  ##  get slot for req if available
    if slot == None: return                       ## #   (p.34)  ### added
                                                  #   slot = (sent_num + 1) % window_size;  ### redundant; what if None?
    sst[my_rank].slots[slot].buf = req            ##  # (p.34) the application writes the message contents in the buffer
    sst[my_rank].slots[slot].index += 1           #   sst[my_rank].slots[slot].index++;
    write_sst(my_rank, "slots", sst[my_rank].slots[slot], slot)  ## wrote to SST just above, but need to multicast
    sent_num += 1                                 #   sent_num++; }
  
  # (p.34) A.2.4  Receiving.  ## if there is a req msg in next slot, increment received_num and call recv
  def receive_req():             # always {  ### made function and called in run
    for i in range(n):           #   for i in 1 to n {  ### fixed to be 0..n-1
      slot = (sst[my_rank].received_num[i]+1) % window_size
                                 #    slot = (sst[my_rank].received_num[i]+1)%window_size
      if sst[i].slots[slot].index == (sst[my_rank].received_num[i]+1) // window_size + 1:
                                 #    if (sst[i].slots[slot].index == (sst[my_rank].received_num[i]+1)/window_size+1) {
        write_sst(my_rank, "received_num", sst[my_rank].received_num[i]+1, i)
                                 #      ++sst[my_rank].received_num[i];
        recv(M(i, sst[my_rank].received_num[i]), i,  sst[my_rank].received_num[i])
                                 #      recv(M(i, sst[my_rank].received_num[i])); }}}


  # (p.34) A.3  Atomic Multicast Delivery in the Steady State
  # A.3.1  Receive.  ## received request msg, update msgs and related indices, but first send null msgs if needed
  def recv(req, i, k):    # on recv(M(i,k)) {
    ### the if-block below is added to avoid stalls by slow senders in delivery of received message,
    ### pseudocode here as in email from Sagar Jha to Vishnu Paladugu on 11/29/19, quoting an email by him dated 7/29/18
                                          # if (I am a sender && this subgroup is not in unordered mode) {  ### ignored
    if my_rank < i and k > sent_num:      #   if (my_rank < i && I have not sent M(my_rank, k)) {  ### used sent_num
      for _ in range(k - sent_num):       ##  for every missing msg:  ### do all at once to be more efficient
        output("sending no-op, case 1")
        send_req(None)                    #     send a null message }
    elif my_rank > i and k-1 > sent_num:  # else if (my_rank > i && I have not sent M(my_rank, k-1)) { ### used sent_num
      for _ in range(k-1 - sent_num):     ##  for every missing msg:  ### do all at once to be more efficient
        output("sending no-op, case 2")
        send_req(None)                    #     send a null message }}
    msgs[gi(i, k)] = req  #   msgs[gi(M(i,k))] = M(i,k);
    write_sst(my_rank, "latest_received_index", k, i)
                          #   sst[my_rank].latest_received_index[i] = k;
    min_index_received, lagging_node_rank = min_and_idx(sst[my_rank].latest_received_index)
                          #   (min_index_received, lagging_node_rank) =
                          #       (min,argmin) i sst[my_rank].latest_received_index[i];
    write_sst(my_rank, "global_index", (min_index_received + 1) * len(G) + lagging_node_rank - 1)
                          #   sst[my_rank].global_index = (min_index_received + 1) * |G| + lagging_node_rank - 1;
                          # }

  # (p.34) A.3.2  Stability and Delivery.
  ## deliver consecutive msgs that have been received by all nodes, in order of global index of the msgs
  def stability_delivery():                 # always {  ### made function and called in run
    stable_msg_idx = Min("global_index")    #   stable_msg_index = Min{sst[*].global_index}

    sorted_keys = sorted(msgs.keys())       ##  sort because msgs must be delivered in increasing global index
    for g_idx in sorted_keys:               #   for (msg : msgs) {
      if g_idx <= stable_msg_idx:           #     if (msg.global_index <= min_stable_msg_index) {  ### min_ deleted
        deliver_upcall(g_idx, msgs[g_idx])  #       deliver_upcall(msg);  ### add first argument, to see global_index in output
        del msgs[g_idx]                     #       msgs.remove(msg.global_index); }}
    if sorted_keys and min(sorted_keys) <= stable_msg_idx:
                                            ##  if there stored reqs are less than received reqs
                                                ### added test to not multicast unnecessarily as this is in an always
      write_sst(my_rank, "latest_delivered_index", stable_msg_idx)
                                            #   sst[my_rank].latest_delivered_index = stable_msg_index }
\end{lstlisting}
\Vex{-2.5}
\caption{Specification of steady-state execution.}
\label{fig-steady}
\end{figure*}

The first part consists of functions \co{send\_req} and \co{get\_buffer}. 
Upon receiving a request message from a client, the \co{receive} handler (lines 1-8) is executed. 
We added a check to avoid processing duplicate requests (lines 4-7). 
Subsequently, function \co{send\_req} is called, which uses function \co{get\_buffer} (line 23) to obtain a slot in field \co{slots} to hold the request. 
Function \co{get\_buffer} returns a pointer to the ring buffer if the request message 
previously in that slot has been successfully received on all the nodes in the group, and \co{nullptr} otherwise. 
However, in our specification, we return the slot number if the reservation is successful and \co{None} otherwise (lines 16 and 19).
The message is then written in all nodes by calling \co{write\_sst} (line 28). 

The second part uses functions \co{receive\_req} and \co{recv} to receive incoming messages from other nodes. 
Function \co{receive\_req} checks for new messages from other nodes in the group (lines 31-40).  
It is run continually to simulate "always" in the pseudocode, by using a nondeterministic random choice to select a function to run in node's \co{run} function's top-level infinite loop. 
Upon receiving a message, function \co{recv} is called, which stores the message in a 
dictionary data structure 
along with calculating \co{global\_index}, 
which represents the highest global index of the message 
that can be safely 
delivered based on the local 
computation (lines 43-63).

The third part uses \co{stability\_delivery} to deliver messages, in order of their global indexes, 
that have been received on all nodes  (lines 66-77).
The minimum of the \co{global\_index} across nodes, known as the \co{stable\_msg\_idx} (line 67), is the index 
up to which messages 
can be safely delivered.



\begin{figure*}
\begin{lstlisting}[firstline=529,lastline=540]
  def deliver_upcall(global_idx, req):              ## execute request req, at decided global index
    output("in deliver_upcall(), gi: ", global_idx, " and req: ", req)    
    if req is None: return                          ## if request is a null msg for no-op, return
    (client, req_id, _) = req                       ## request is form (client, req_id, cmd)
    if not some(sent(('response', _req_id, _), to= _client)):  ## if request has not been responded to before
      state, res = execute(global_idx, req, state)  ## execute request at global_idx in state
      send(('response', req_id, res), to= client)   ## send response to client
      output("response sent to the client/sim process, index: ", global_idx, "response: ", res)

  def execute(global_idx, req, state):              ## execute the command in req in given state
    (_, req_id, _) = req                            ## request is of form (client, req_id, cmd)
    return (state+[(global_idx, req)], req_id)      ## return call history and req_id as new state and result
\end{lstlisting}
\Vex{-2}
\caption{Specification of functions for delivering and executing requests.} 
\label{fig-deliver-upcall}
\end{figure*}

The pseudocode lacks a definition of function \co{deliver\_upcall}. We added it, as shown in Fig.~\ref{fig-deliver-upcall}.
In our specification, we have abstracted the atomic multicast delivery of a request message from its execution.
Function \co{deliver\_upcall} takes \co{global\_index} and \co{request} and signifies the delivery of the request (lines 1-8). 
If a corresponding response has not been sent for a request,
it calls the function \co{execute} (lines 9-11) to signify execution of the request.
This approach helps prevent execution of duplicate requests that may arise due to client resends.

\section{Specifying view change} 

Upon encountering the failure of a node, the group undergoes a membership change 
by employing a two-phase commit. 
A key part of the algorithm is leader selection. Fig.~\ref{fig-leader-selection} shows its complete precise specification in DistAlgo.




The leader-selection algorithm is specified by two functions, \co{find\_new\_leader} (lines 1-4) and an "always" running function \acm{\linebreak} \co{leader\_selection} (lines 6-25).
Function \co{find\_new\_leader} selects the first non-suspected node as the leader. 
In function \co{leader\_selection}, if a new leader selected is different from the current leader (line 8), 
it waits until all non-suspected nodes recognize it as the leader before continuing (lines 12-23).



\begin{figure*}
\begin{lstlisting}[firstline=376,lastline=402]
  def find_new_leader(r):  # find_new_leader(r) {
    for i in range(len(curr_view.members)):  #   for (int i = 0; i < curr_view.max_rank; ++i) {  ### max_rank replaced
      if sst[r].suspected[i]: continue       #     if (sst[r].suspected[i]) continue;
      else: return i                         #     else return i }}

  # (p.35)  ## update the current view, at the end, with the new leader
  def leader_selection():                           # always {  ### made function and called in run
    new_leader = find_new_leader(my_rank)           # new_leader = find_new_leader(my_rank)
    if new_leader != curr_view.leader_rank:         # if (new_leader != curr_view.leader_rank && new_leader == my_rank)
      if new_leader == my_rank:                     ### split 2 conjuncts, to add the else-branch for the second
      # all_others_agree = True                     #   bool all_others_agree = true  ### moved into while-loop 
      ### if not moved, if it becomes False in for-loop below, it stays False, and the while-loop never stops
        while find_new_leader(my_rank) == my_rank:  #   while (find_new_leader(my_rank) == my_rank) {
          --receive_messages                        ##    yield to receive msgs
          ### needed to receive updates to SST which may result in new leader selection  ### break atomicity
        
          all_others_agree = True                   ###   moved here from outside while-loop, as explained above
          for r in range(len(sst)):                 #     for (r: SST.rows) {
            if not sst[my_rank].suspected[r]:       #       if (sst[my_row].suspected[r] == false)
              all_others_agree = all_others_agree and (find_new_leader(r) == my_rank)
                                                    #         all_others_agree &&= (find_new_leader(r) == my_rank) }
          if all_others_agree:                      #     if (all_others_agree) {
            curr_view.leader_rank = my_rank         #       curr_view.leader_rank = my_rank;
            output("I am the new leader!!!")
            break                                   #       break; }}}
      else:                                         ##   else:  ### added else-branch, for when new leader is not self
        curr_view.leader_rank = new_leader          ##     set current view's leader to be new leader
\end{lstlisting}
\Vex{-2}
\caption{Specification of leader selection.}
\label{fig-leader-selection}
\end{figure*}


\section{Runtime checking and analysis}

\subsection{Manual inspection and automated checking}

To help ensure the correctness of the protocol specification, %
we perform careful manual inspection 
automated testing, and increasingly systematic runtime checking of safety and progress properties,
and repeat this process for each anomaly and improvement discovered until all inspections, tests, and checks pass.
This approach led to a complete precise specification in DistAlgo,
after filling in missing details in the English and pseudocode description and resolving additional issues.

The testing and checking methodology 
consists of configuring and executing the protocol with varying numbers of requests and member nodes, ring buffer size, etc., as well as introducing random node failures. 
The systematic runtime checking was enabled by a general framework in DistAlgo
for runtime checking of safety and liveness properties without touching the complete protocol specification~\cite{RV2020}.

\subsection{Properties checked}





%
%
%

\noindent For property checking, the following messages are used: 

\begin{itemize}
    \item \co{p.sent(`deliver\_upcall', i, req, t)} for \co{p} calling \acm{\newline} \co{deliver\_upcall(i, req)} at time \co{t}
    
    \item \co{p.sent(`execute', i, req)} for \co{p} calling 
    \co{execute(i, req)} 
    
     \item \co{p.receive('request', req)} for message \co{('request', req)} \acm{\newline} received by \co{p} 
\end{itemize}

An important property discussed in the paper \cite{Jha2019DerechoFS} is the round-robin message delivery.

\mysubpar{Delivery ordering} 
"Derecho uses a simple round-robin delivery order: Each active sender can provide one multicast per delivery cycle, and the messages are delivered in round-robin manner. The global index of M(i, k), gi(M(i, k)) is the position of this message in the round-robin ordering."



\label{sec: delivery_ordering_check}
\begin{lstlisting}[%caption=Delivery Ordering, 
label={lst: delivery_ordering_check}, numbers=none]
    each p.sent('deliver_upcall', i1, req1, t),
         p.sent('deliver_upcall', i2, req2, t2)
         has (not i2>i1 or t2>t)
\end{lstlisting}


We also check the following well-known properties, 
taken and quoted exactly from Paxos-SB \cite{paxos-sb}, except that an update in Paxos-SB corresponds to a client request, and a server is a node process 
in Derecho.

\mysubpar{Validity} "Only an update that was introduced by a client (and subsequently initiated by a server) may be executed."

%
\label{sec: validity_check}
\begin{lstlisting}[label={lst: validity_check}, numbers=none]
    each p.sent('execute', i, req)
         has some p1.received('request', _req)
\end{lstlisting}


\mysubpar{Agreement} "If two servers execute the i\textsuperscript{th} update, then these updates are identical."



\label{sec: agreement_check}
\begin{lstlisting}[label={lst: agreement_check}, numbers=none]
    each p1.sent('execute', i, req1), 
         p2.sent('execute', i, req2) 
         has req1=req2
\end{lstlisting} 


\mysubpar{Uniform integrity}
"If a server executes an update on sequence number i, then the server does not execute the update on any other sequence number i' $>$ i."


\label{sec: uniform_integrity_check}
\begin{lstlisting}[%caption=Uniform Integrity, 
label={lst: uniform_integrity_check}, numbers=none]
    each p.sent('execute', i, req) has
         not some =p.sent('execute', i2, =req) has i2>i
\end{lstlisting}

Additionally, we use aggregation queries to specify and check important progress properties, e.g., the total number of
executed request equals the total number of client requests.
\begin{lstlisting}[numbers=none]]
    count(req: p.receive('request', req)
    = count(req: p.sent('execute', i, req)
\end{lstlisting}

\subsection{Issues found and fixed}

Our specification and checking approach---by following the pseudocode exactly, facilitated by the already detailed pseudocode of Derecho---has also led to finding and fixing some issues in the pseudocode.
Many of these issues were difficult to identify in the original pseudocode
due to usual problems with pseudocode, compounded with the complexity of Derecho,
but became evident after the specification in DistAlgo.
While most of the issues were easy to find and fix, others were not.

Some initial issues (e.g., typos) and more were already addressed in an errata~\cite[Errata, page 50]{Jha2019DerechoFS-Cornell} and a dissertation~\cite{Jha22thesis}, and some others (e.g., adding null messages to prevent stalls, Fig.~\ref{fig-steady} lines 47-54) 
were resolved with help from Derecho developers~\cite{Jha19email,Jha23email}.
In all cases, Derecho developers have checked and confirmed that these bugs are not in 
their implementation in C++.

The first bugs~\cite[Errata, page 50]{Jha2019DerechoFS-Cornell} were found mostly by manual inspection, when
writing and examining the DistAlgo specification by following the text
description of the logic and the pseudocode of the protocol and cross
checking.
The rest were essentially all discovered first by automated testing and
checking and then by manual inspection, adding tests and checks, and
running again to confirm. The fixes proposed passed all tests, checks, and inspections.

We discuss two examples issues discovered and fixed, for steady-state execution and view change, respectively.
Despite being minor in hindsight, such issues were tricky to discover due to complex control flows from high nondeterminism.
Both helped improve Derecho's pseudocode~\cite[pages 144 and 149]{Jha22thesis}.


\mypar{Overwriting in ring buffer}
%
In steady-state execution,
field \co{slots} in SST stores requests in a ring buffer
of size \co{window\_size} (Fig.~\ref{fig-system-state}, line 13).
To prevent overwriting a slot,
\co{get\_buffer} should return \co{nullptr} if
the number of pending messages 
equals the buffer size, 
i.e., \co{sent\_num} (number of messages sent by the node) - \co{completed\_num} (number of sent messages that have been received by all nodes) \m{=} \co{window\_size} (Fig.~\ref{fig-steady} line 13).
Instead, the check
uses "\m{<}"  originally~\cite[Sec. A.2.3, page 33]{Jha2019DerechoFS} and "\m{>}" in a later errata \cite[Errata, page 50]{Jha2019DerechoFS-Cornell}). 

Note that the improved Derecho pseudocode uses \m{\geq}~\cite[page 144]{Jha22thesis}, which is also correct and was how we first proposed to fix.  Using "\m{=}" is more precise, because for the protocol to be correct, "\m{>}" should never happen.

This is clearly a minor bug, and finding the "fix" in the errata was relatively quick.
However, discovering more issues after that and debugging them were highly involved, but solved with help by both manual inspection and automatic checking.
The bug manifested as a deadlock in function \co{receive\_req} (Fig.~\ref{fig-steady}, line 35) for nodes that did not receive the previous message in the slot.
It only happens when the ring buffer is full, after a long execution trace of member nodes handling many client requests, even for a small buffer size.


\mypar{Deadlock in leader selection}
%
Leader selection (Fig.~\ref{fig-leader-selection}) selects 
the first non-suspected node (lines 1-4) as the new leader and uses variable \co{all\_others\_agree} to track a logical conjunction checking if all nodes agree with the new leader (lines 10, 18, and 20).
We discovered that if a node did not initially agree, i.e., \co{find\_new\_leader} returned a different leader, \co{all\_others\_agree} would be set to \co{False} (line 18) on the new leader's node, and cause a deadlock because \co{all\_others\_agree} can never be set to \co{True} again.

To fix this, we (1) move the first write to \co{all\_others\_agree} from outside (line 10) to inside (line 15) of the \co{while}, to reset it in each iteration, and (2) add an \co{else} branch to update the new leader for non-leader nodes (lines 24-25).

\subsection{Resulting specification and direct execution}

Table \ref{table:derecho_loc} shows the size of specification of Derecho in DistAlgo.
\begin{table}[h]
\centering
\caption{Specification size (in number of lines, including output lines, excluding empty or 
comment-only lines) 
for Derecho specification in DistAlgo  (\co{derecho.da} in Appendix~\ref{sec-derecho.da} excluding method \co{main} and class \co{Sim}).}
\acm{\Vex{-2.5}}
\begin{tabular}{l | r}
Protocol component & Size\\\hline
state and helper functions& 95\\
steady-state execution, incl.\ delivering\&executing reqs & 63\\
view change & 132\\
\co{import}s, helper, choices in \co{run} & 14\\\hline
total & 304\\\hline 
\end{tabular}
\label{table:derecho_loc}
\end{table}



DistAlgo specifications can be run directly. For example, 
 Derecho specification in a file \co{derecho.da} (Appendix \ref{sec-derecho.da}) can be executed with Python 3.7 by simply running \arxiv{\linebreak} \co{pip install pyDistAlgo} to install DistAlgo and then running \co{python -m da derecho.da}.
To run the system with a failure of a node injected, set \co{test\_failure} to \co{True} in method \co{main}. 


The specification with steady-state and failure-induced membership change runs smoothly.
For instance, with 1000 requests, three nodes, one client, window-size 400 and no failure, the average time over 10 runs was 4.8 seconds, measured on a 2.6 GHz 6-Core Intel Core i7 CPU with 16 GB RAM running macOS Ventura and Python 3.7.12. A complete sample run for three nodes, one client, and ten requests, with window-size as ten can be found in Appendix \ref{sec: derecho output}.

Note that this is essentially the same speed for runtime checking.
Our specification is not currently optimized for efficiency;
it directly runs many reducer functions 
aggregations, sorting, and output constantly, each taking linear time or worse.
These high-level functions are essential for ensuring correctness. 
The efficiency of such expensive functions can be improved drastically, asymptotically, by using incrementalization, as in~\cite{Liu+17DistPL-TOPLAS}.
Our goal in this work is to develop a complete, precise, and correct specification.


\section{Related work and conclusion}
\label{sec-related}


Significant efforts have been devoted to specifying, testing, analyzing, checking, and verifying distributed algorithms, as evidenced by works such as \cite{IronFleet,CLS16,PLSS17}. 
Furthermore, various specification languages and verification tools, such as TLA+ Toolbox \cite{Lam94, Lam02, Mic20} and Ivy \cite{MP20}, have been developed to aid in this task. 
Another area of focus is verifying the executable specifications of distributed consensus protocols, as demonstrated by projects like IronFleet \cite{IronFleet} and Verdi \cite{Verdi}. 
For example, Verdi \cite{Verdi-Raft} provides a verified implementation of Raft in Coq \cite{Coq} with 50,000 lines of proof. 
Nonetheless, the challenge remains in the significant development efforts and expertise required for such verification.

Runtime verification (RV) is a useful tool for ensuring the correct functioning of complex distributed algorithms and their executable specifications, especially for real-world applications written in general-purpose programming languages, where manual verification can be prohibitively difficult. 
Several RV tools have been developed, including MoP \cite{Chen2005JavaMOPAM,mop2007} and ELarva \cite{ELarva}. However, they either have not been applied to general distributed systems or have more complex property specification and checking processes. 
Although WidsChecker \cite{WiDSChecker} found bugs in Paxos' IO automata specification, it is tightly coupled with programs developed using WidsToolkit, is not publicly available, and requires complex specification scripts. 
In contrast, our runtime verification framework \cite{RV2020} enables clear and precise high-level specifications of properties, allowing for the detection of subtle bugs in protocol pseudocode and earlier specifications. 



Previous research has explored different approaches for producing executable code from formal specifications, such as from process algebras \cite{han2001}, I/O automata \cite{Georgiou2009AutomatedIO}, and various high-level languages, like Dedalus \cite{Dedalus2011}, Bloom \cite{Alvaro2011ConsistencyAI}, EventML \cite{eventML}, and DAHL \cite{lopes_navarro_rybalchenko_singh_2010}. 
However, DistAlgo~\cite{Liu+12DistPL-OOPSLA,Liu+17DistPL-TOPLAS} stands out as a language specifically designed for easy and precise expression and direct execution of distributed algorithms. 
Additional work using DistAlgo includes many distributed algorithms specified in DistAlgo, e.g.,~\cite{Liu+17DistPL-TOPLAS} and 
on github (\url{https://github.com/search?q=distalgo}),
automatic transformations of Event-B models \cite{EventB_DistAlgo} into DistAlgo, as well as uses in 
various BS and MS theses, e.g., \cite{widmer2020byzantine, lazic2021library, Shi22thesis}.



For our study, we selected Derecho due to its importance in various high-speed data replication in intelligent IoT edge applications \cite{CascadeApplied22}, with optimization techniques like Spindle \cite{Spindle2022} improving bandwidth utilization for small messages.
Nevertheless, compared to multi-Paxos \cite{vra-2015}, with a corresponding executable specification in DistAlgo~\cite{liu-ppdp}, Derecho is much more complex with many more pieces of information maintained in much more sophisticated control flows and with less complete pseudocode. This made it challenging to determine some of the missing details in the specification.

Our directly executable specification closely corresponds to the protocol pseudocode, and this helped us better understand the protocol, leading us to identify missing details required for a complete, precise, executable specification.
We use runtime checking to check important safety and progress properties of the protocol.
Future work includes the incorporation of various fault-injection testing methods with the current runtime verification framework, use of the specification in DistAlgo to help with proof development~\cite{DerechoDDS} for both safety and liveness,
as well as automated ways to correlate formal specifications in DistAlgo with efficient implementations in lower-level languages such as C++.


\subsection*{Acknowledgments} 
We thank the Derecho team, Ken Birman and Sagar Jha in particular, for their prompt replies to our inquiries and their detailed and helpful explanations about the Derecho protocol.
We thank Thejesh Arumalla for careful study of work by the Derecho team and greatly helpful questions and discussions.
This work was supported in part by NSF under grant 
CCF-1954837 
and ONR under grant
N00014-21-1-2719. 

\bibliographystyle{alpha}
\bibliography{bibs}

\newcommand{\etalchar}[1]{$^{#1}$}
\begin{thebibliography}{WWA{\etalchar{+}}16}

\bibitem[ACHM11]{Alvaro2011ConsistencyAI}
Peter Alvaro, Neil Conway, Joseph~M. Hellerstein, and William~R. Marczak.
\newblock Consistency analysis in bloom: a calm and collected approach.
\newblock In {\em Conference on Innovative Data Systems Research}, 2011.

\bibitem[AMC{\etalchar{+}}11]{Dedalus2011}
Peter Alvaro, William~R. Marczak, Neil Conway, Joseph~M. Hellerstein, David
  Maier, and Russell Sears.
\newblock Dedalus: Datalog in time and space.
\newblock In Oege de~Moor, Georg Gottlob, Tim Furche, and Andrew Sellers,
  editors, {\em Datalog Reloaded}, pages 262--281, Berlin, Heidelberg, 2011.
  Springer Berlin Heidelberg.

\bibitem[Bic09]{eventML}
Mark Bickford.
\newblock Component specification using event classes.
\newblock In Grace~A. Lewis, Iman Poernomo, and Christine Hofmeister, editors,
  {\em Component-Based Software Engineering}, pages 140--155, Berlin,
  Heidelberg, 2009. Springer Berlin Heidelberg.

\bibitem[BJ87]{birman1987exploiting}
Kenneth~P. Birman and Thomas~A. Joseph.
\newblock Exploiting virtual synchrony in distributed systems.
\newblock {\em SIGOPS Oper. Syst. Rev.}, 21(5):123--138, November 1987.

\bibitem[CFG12]{ELarva}
Christian Colombo, Adrian Francalanza, and Rudolph Gatt.
\newblock Elarva: A monitoring tool for {Erlang}.
\newblock In Sarfraz Khurshid and Koushik Sen, editors, {\em Runtime
  Verification}, pages 370--374, Berlin, Heidelberg, 2012. Springer Berlin
  Heidelberg.

\bibitem[CL21]{ChaLiu20Liveness-PODC21}
Saksham Chand and Yanhong~A. Liu.
\newblock Brief announcement: What's live? understanding distributed consensus.
\newblock pages 565--568, July 2021.

\bibitem[CLS16]{CLS16}
Saksham Chand, Yanhong~A. Liu, and Scott~D. Stoller.
\newblock Formal verification of multi-paxos for distributed consensus.
\newblock In John Fitzgerald, Constance Heitmeyer, Stefania Gnesi, and Anna
  Philippou, editors, {\em FM 2016: Formal Methods}, pages 119--136, Cham,
  2016. Springer International Publishing.

\bibitem[Coq]{Coq}
Coq, a formal proof management system. https://coq.inria.fr/.

\bibitem[CR05]{Chen2005JavaMOPAM}
Feng Chen and Grigore Roşu.
\newblock Java-mop: A monitoring oriented programming environment for java.
\newblock In {\em International Conference on Tools and Algorithms for
  Construction and Analysis of Systems}, 2005.

\bibitem[CR07]{mop2007}
Feng Chen and Grigore Ro\c{s}u.
\newblock Mop: An efficient and generic runtime verification framework.
\newblock In {\em Proceedings of the 22nd Annual ACM SIGPLAN Conference on
  Object-Oriented Programming Systems, Languages and Applications}, OOPSLA '07,
  page 569–588, New York, NY, USA, 2007. Association for Computing Machinery.

\bibitem[der]{derecho-da-github}
Derecho distalgo github repository.
  https://github.com/unicomputing/derecho-distalgo.

\bibitem[GLMT09]{Georgiou2009AutomatedIO}
Chryssis Georgiou, Nancy~A. Lynch, Panayiotis Mavrommatis, and Joshua~A.
  Tauber.
\newblock Automated implementation of complex distributed algorithms specified
  in the ioa language.
\newblock {\em International Journal on Software Tools for Technology
  Transfer}, 11:153--171, 2009.

\bibitem[Gra20]{EventB_DistAlgo}
Alexis Grall.
\newblock Automatic generation of distalgo programs from event-b models.
\newblock In Alexander Raschke, Dominique M{\'e}ry, and Frank Houdek, editors,
  {\em Rigorous State-Based Methods}, pages 414--417, Cham, 2020. Springer
  International Publishing.

\bibitem[HCS01]{han2001}
D.~Hansel, R.~Cleaveland, and S.A. Smolka.
\newblock Distributed prototyping from validated specifications.
\newblock In {\em Proceedings 12th International Workshop on Rapid System
  Prototyping. RSP 2001}, pages 97--102, 2001.

\bibitem[HHK{\etalchar{+}}15]{IronFleet}
Chris Hawblitzel, Jon Howell, Manos Kapritsos, Jacob~R. Lorch, Bryan Parno,
  Michael~L. Roberts, Srinath Setty, and Brian Zill.
\newblock Ironfleet: Proving practical distributed systems correct.
\newblock In {\em Proceedings of the 25th Symposium on Operating Systems
  Principles}, SOSP '15, page 1–17, New York, NY, USA, 2015. Association for
  Computing Machinery.

\bibitem[JBG{\etalchar{+}}19a]{Jha2019DerechoFS}
Sagar Jha, Jonathan Behrens, Theo Gkountouvas, Matthew Milano, Weijia Song,
  Edward Tremel, Robbert~Van Renesse, Sydney Zink, and Kenneth~P. Birman.
\newblock Derecho: Fast state machine replication for cloud services.
\newblock {\em ACM Trans. Comput. Syst.}, 36(2), April 2019.

\bibitem[JBG{\etalchar{+}}19b]{Jha2019DerechoFS-Cornell}
Sagar Jha, Jonathan Behrens, Theo Gkountouvas, Matthew Milano, Weijia Song,
  Edward Tremel, Robbert van Renesse, Sydney Zink, and Kenneth~P. Birman.
\newblock Derecho: Fast state machine replication for cloud services.
\newblock {\em ACM Trans. Comput. Syst.}, 36:4:1--4:49, 2019.
\newblock with Errata, page 50, Nov. 2019.
  \url{https://www.cs.cornell.edu/ken/derecho-tocs.pdf}.

\bibitem[Jha19]{Jha19email}
Sagar Jha.
\newblock Re: Null sends, November 27 2019.
\newblock Email, with Vishnu Paladugu, forwarding an email dated Jul 29, 2018.

\bibitem[Jha22]{Jha22thesis}
Sagar Jha.
\newblock {\em RDMA-accelerated state machine for cloud services}.
\newblock PhD thesis, Cornell University, Ithaca, NY, 12 2022.
\newblock
  \url{https://www.cs.cornell.edu/projects/Quicksilver/public_pdfs/dissertation.pdf}.

\bibitem[Jha23]{Jha23email}
Sagar Jha.
\newblock Re: Understanding the derecho's view-change algorithm, April 30 2023.
\newblock Email with Kumar Shivam.

\bibitem[JRB22]{Spindle2022}
Sagar Jha, Lorenzo Rosa, and Ken Birman.
\newblock Spindle: Techniques for optimizing atomic multicast on rdma.
\newblock In {\em 2022 IEEE 42nd International Conference on Distributed
  Computing Systems (ICDCS)}, pages 1085--1097, 2022.

\bibitem[KA08]{paxos-sb}
Jonathan Kirsch and Yair Amir.
\newblock Paxos for system builders: An overview.
\newblock In {\em Proceedings of the 2nd Workshop on Large-Scale Distributed
  Systems and Middleware}, LADIS '08, New York, NY, USA, 2008. Association for
  Computing Machinery.

\bibitem[Lam94]{Lam94}
Leslie Lamport.
\newblock The temporal logic of actions.
\newblock {\em ACM Trans. Program. Lang. Syst.}, 16(3):872–923, may 1994.

\bibitem[Lam98]{partTimeParliament}
Leslie Lamport.
\newblock The part-time parliament.
\newblock {\em ACM Trans. Comput. Syst.}, 16(2):133–169, may 1998.

\bibitem[Lam02]{Lam02}
Leslie Lamport.
\newblock {\em Specifying Systems: The TLA+ Language and Tools for Hardware and
  Software Engineers}.
\newblock Addison-Wesley Longman Publishing Co., Inc., USA, 2002.

\bibitem[Laz21]{lazic2021library}
Aleksandar Lazic.
\newblock The library of distributed protocols.
\newblock Master's thesis, University of Fribourg, 2021.

\bibitem[LCS19]{liu-ppdp}
Yanhong~A. Liu, Saksham Chand, and Scott~D. Stoller.
\newblock Moderately complex paxos made simple: High-level executable
  specification of distributed algorithms.
\newblock In {\em Proceedings of the 21st International Symposium on Principles
  and Practice of Declarative Programming}, PPDP '19, New York, NY, USA, 2019.
  Association for Computing Machinery.

\bibitem[LLPZ07]{WiDSChecker}
Xuezheng Liu, Wei Lin, Aimin Pan, and Zheng Zhang.
\newblock {WiDS} checker: Combating bugs in distributed systems.
\newblock In {\em 4th USENIX Symposium on Networked Systems Design \&
  Implementation (NSDI 07)}, Cambridge, MA, April 2007. USENIX Association.

\bibitem[LLS17]{Distalgo17lang}
Yanhong~A. Liu, Bo~Lin, and Scott Stoller.
\newblock {{DistAlgo} Language Description}.
\newblock \url{http://distalgo.cs.stonybrook.edu}, March 2017.

\bibitem[LNRS10]{lopes_navarro_rybalchenko_singh_2010}
NUNO~P. LOPES, JUAN~A. NAVARRO, ANDREY RYBALCHENKO, and ATUL SINGH.
\newblock Applying prolog to develop distributed systems.
\newblock {\em Theory and Practice of Logic Programming}, 10(4-6):691–707,
  2010.

\bibitem[LS20]{RV2020}
Yanhong~A. Liu and Scott~D. Stoller.
\newblock Assurance of distributed algorithms and systems: Runtime checking of
  safety and liveness.
\newblock In Jyotirmoy Deshmukh and Dejan Ni{\v{c}}kovi{\'{c}}, editors, {\em
  Runtime Verification}, pages 47--66, Cham, 2020. Springer International
  Publishing.

\bibitem[LSL17]{Liu+17DistPL-TOPLAS}
Yanhong~A. Liu, Scott~D. Stoller, and Bo~Lin.
\newblock From clarity to efficiency for distributed algorithms.
\newblock {\em ACM Transactions on Programming Languages and Systems},
  39(3):12:1--12:41, May 2017.

\bibitem[LSLG12]{Liu+12DistPL-OOPSLA}
Yanhong~A. Liu, Scott~D. Stoller, Bo~Lin, and Michael Gorbovitski.
\newblock From clarity to efficiency for distributed algorithms.
\newblock In {\em Proceedings of the 27th ACM SIGPLAN Conference on
  Object-Oriented Programming, Systems, Languages and Applications}, pages
  395--410, 2012.

\bibitem[Mic]{Mic20}
Microsoft research. the tla toolbox.
  http://lamport.azurewebsites.net/tla/toolbox.html.

\bibitem[MP20]{MP20}
Kenneth~L. McMillan and Oded Padon.
\newblock Ivy: A multi-modal verification tool for distributed algorithms.
\newblock In {\em Computer Aided Verification: 32nd International Conference,
  CAV 2020, Los Angeles, CA, USA, July 21–24, 2020, Proceedings, Part II},
  page 190–202, Berlin, Heidelberg, 2020. Springer-Verlag.

\bibitem[OL88]{VR}
Brian~M. Oki and Barbara~H. Liskov.
\newblock Viewstamped replication: A new primary copy method to support
  highly-available distributed systems.
\newblock In {\em Proceedings of the Seventh Annual ACM Symposium on Principles
  of Distributed Computing}, PODC '88, page 8–17, New York, NY, USA, 1988.
  Association for Computing Machinery.

\bibitem[PLSS17]{PLSS17}
Oded Padon, Giuliano Losa, Mooly Sagiv, and Sharon Shoham.
\newblock Paxos made epr: Decidable reasoning about distributed protocols.
\newblock {\em Proc. ACM Program. Lang.}, 1(OOPSLA), oct 2017.

\bibitem[RJB21]{DerechoDDS}
Lorenzo Rosa, Sagar Jha, and Ken Birman.
\newblock {DerechoDDS: Efficiently leveraging RDMA for fast and consistent data
  distribution}.
\newblock In {\em {CARS 2021 6th International Workshop on Critical Automotive
  Applications: Robustness \& Safety}}, M{\"u}nich, Germany, September 2021.

\bibitem[Shi22]{Shi22thesis}
Kumar Shivam.
\newblock Specification and runtime checking of algorithms for replication and
  consensus in distributed systems.
\newblock Master's thesis, Stony Brook University, 2022.

\bibitem[SYL{\etalchar{+}}22]{CascadeApplied22}
Weijia Song, Yuting Yang, Thompson Liu, Andrea Merlina, Thiago Garrett, Roman
  Vitenberg, Lorenzo Rosa, Aahil Awatramani, Zheng Wang, and Ken Birman.
\newblock Cascade: An edge computing platform for real-time machine
  intelligence.
\newblock In {\em Proceedings of the 2022 Workshop on Advanced Tools,
  Programming Languages, and PLatforms for Implementing and Evaluating
  Algorithms for Distributed Systems}, ApPLIED '22, page 2–6, New York, NY,
  USA, 2022. Association for Computing Machinery.

\bibitem[VRA15]{vra-2015}
Robbert Van~Renesse and Deniz Altinbuken.
\newblock Paxos made moderately complex.
\newblock {\em ACM Comput. Surv.}, 47(3), feb 2015.

\bibitem[Wid20]{widmer2020byzantine}
Roland Widmer.
\newblock Byzantine-fault tolerant algorithms in {DistAlgo}. bachelors thesis,
  2020.

\bibitem[WWA{\etalchar{+}}16]{Verdi-Raft}
Doug Woos, James~R. Wilcox, Steve Anton, Zachary Tatlock, Michael~D. Ernst, and
  Thomas Anderson.
\newblock Planning for change in a formal verification of the raft consensus
  protocol.
\newblock In {\em Proceedings of the 5th ACM SIGPLAN Conference on Certified
  Programs and Proofs}, CPP 2016, page 154–165, New York, NY, USA, 2016.
  Association for Computing Machinery.

\bibitem[WWP{\etalchar{+}}15]{Verdi}
James~R. Wilcox, Doug Woos, Pavel Panchekha, Zachary Tatlock, Xi~Wang,
  Michael~D. Ernst, and Thomas Anderson.
\newblock Verdi: A framework for implementing and formally verifying
  distributed systems.
\newblock {\em SIGPLAN Not.}, 50(6):357–368, jun 2015.

\end{thebibliography}

\appendix
\onecolumn
\section{Derecho executable specification in DistAlgo}
\label{sec-derecho.da}
\lstinputlisting[caption=Complete specification in DistAlgo.
]{derecho_code.da}


\section{Derecho sample output} 
\label{sec: derecho output}
\begin{lstlisting}[style=outputstyle, caption={Derecho output for group consisting of 3 nodes, 1 client, and 10 requests per client with window-size as 10.}]
[190] da.api<MainProcess>:INFO: <Node_:4ec01> initialized at 127.0.0.1:(UdpTransport=13024, TcpTransport=44810).
[190] da.api<MainProcess>:INFO: Starting program <module 'derecho' from './derecho.da'>...
[190] da.api<MainProcess>:INFO: Running iteration 1 ...
[190] da.api<MainProcess>:INFO: Waiting for remaining child processes to terminate...(Press "Ctrl-C" to force kill)
[239] derecho.Node<Node:b4c02>:OUTPUT: initial group:  [<Node:b4c02>, <Node:b4c03>, <Node:b4c04>]
[241] derecho.Node<Node:b4c03>:OUTPUT: initial group:  [<Node:b4c02>, <Node:b4c03>, <Node:b4c04>]
[243] derecho.Node<Node:b4c04>:OUTPUT: initial group:  [<Node:b4c02>, <Node:b4c03>, <Node:b4c04>]
./derecho.da:566: DeprecationWarning: time.clock has been deprecated in Python 3.3 and will be removed from Python 3.8: use time.perf_counter or time.process_time instead
  metrics[tid] = RequestMetrics(time.clock(), time.time())
[263] derecho.Sim<Sim:b4c05>:OUTPUT: sent request:  0  to node:  [<Node:b4c04>]
[286] derecho.Sim<Sim:b4c05>:OUTPUT: sent request:  1  to node:  [<Node:b4c04>]
[286] derecho.Sim<Sim:b4c05>:OUTPUT: sent request:  2  to node:  [<Node:b4c03>]
[292] derecho.Node<Node:b4c04>:OUTPUT: request:  (<Sim:b4c05>, 0, b'd')  received from client:  <Sim:b4c05>
[293] derecho.Node<Node:b4c04>:OUTPUT: request:  (<Sim:b4c05>, 1, b'\xe1')  received from client:  <Sim:b4c05>
[309] derecho.Sim<Sim:b4c05>:OUTPUT: sent request:  3  to node:  [<Node:b4c02>]
[310] derecho.Node<Node:b4c03>:OUTPUT: sending no-op, case 1
[312] derecho.Node<Node:b4c03>:OUTPUT: request:  (<Sim:b4c05>, 2, b'+')  received from client:  <Sim:b4c05>
[315] derecho.Node<Node:b4c02>:OUTPUT: sending no-op, case 1
[319] derecho.Sim<Sim:b4c05>:OUTPUT: sent request:  4  to node:  [<Node:b4c02>]
[319] derecho.Sim<Sim:b4c05>:OUTPUT: sent request:  5  to node:  [<Node:b4c03>]
[319] derecho.Sim<Sim:b4c05>:OUTPUT: sent request:  6  to node:  [<Node:b4c03>]
[320] derecho.Sim<Sim:b4c05>:OUTPUT: sent request:  7  to node:  [<Node:b4c02>]
[320] derecho.Sim<Sim:b4c05>:OUTPUT: sent request:  8  to node:  [<Node:b4c02>]
[320] derecho.Sim<Sim:b4c05>:OUTPUT: sent request:  9  to node:  [<Node:b4c03>]
[320] derecho.Sim<Sim:b4c05>:OUTPUT: done sending all requests
[322] derecho.Node<Node:b4c02>:OUTPUT: sending no-op, case 1
[325] derecho.Node<Node:b4c02>:OUTPUT: request:  (<Sim:b4c05>, 3, b'\x14')  received from client:  <Sim:b4c05>
[325] derecho.Node<Node:b4c02>:OUTPUT: request:  (<Sim:b4c05>, 4, b'\xbf')  received from client:  <Sim:b4c05>
[325] derecho.Node<Node:b4c02>:OUTPUT: request:  (<Sim:b4c05>, 7, b'\x06')  received from client:  <Sim:b4c05>
[326] derecho.Node<Node:b4c02>:OUTPUT: request:  (<Sim:b4c05>, 8, b'\xb3')  received from client:  <Sim:b4c05>
[341] derecho.Node<Node:b4c03>:OUTPUT: request:  (<Sim:b4c05>, 5, b'\x83')  received from client:  <Sim:b4c05>
[342] derecho.Node<Node:b4c03>:OUTPUT: request:  (<Sim:b4c05>, 6, b'\xca')  received from client:  <Sim:b4c05>
[342] derecho.Node<Node:b4c03>:OUTPUT: request:  (<Sim:b4c05>, 9, b'0')  received from client:  <Sim:b4c05>
[344] derecho.Node<Node:b4c04>:OUTPUT: sending no-op, case 2
[345] derecho.Node<Node:b4c03>:OUTPUT: in deliver_upcall(), gi:  0  and req:  None
[345] derecho.Node<Node:b4c03>:OUTPUT: in deliver_upcall(), gi:  1  and req:  None
[346] derecho.Node<Node:b4c03>:OUTPUT: in deliver_upcall(), gi:  2  and req:  (<Sim:b4c05>, 0, b'd')
[346] derecho.Node<Node:b4c03>:OUTPUT: response sent to the client/sim process, index:  2 response:  0
./derecho.da:598: DeprecationWarning: time.clock has been deprecated in Python 3.3 and will be removed from Python 3.8: use time.perf_counter or time.process_time instead
  cur_metric.cpu_end_time = time.clock()
[346] derecho.Sim<Sim:b4c05>:OUTPUT: ---- Received tid:  0  from:  <Node:b4c03>  with num_resp:  1
[347] derecho.Node<Node:b4c04>:OUTPUT: sending no-op, case 2
[351] derecho.Node<Node:b4c03>:OUTPUT: in deliver_upcall(), gi:  3  and req:  None
[351] derecho.Node<Node:b4c03>:OUTPUT: in deliver_upcall(), gi:  4  and req:  (<Sim:b4c05>, 2, b'+')
[351] derecho.Node<Node:b4c04>:OUTPUT: in deliver_upcall(), gi:  0  and req:  None
[351] derecho.Node<Node:b4c03>:OUTPUT: response sent to the client/sim process, index:  4 response:  2
[351] derecho.Sim<Sim:b4c05>:OUTPUT: ---- Received tid:  2  from:  <Node:b4c03>  with num_resp:  2
[351] derecho.Node<Node:b4c04>:OUTPUT: in deliver_upcall(), gi:  1  and req:  None
[351] derecho.Node<Node:b4c04>:OUTPUT: in deliver_upcall(), gi:  2  and req:  (<Sim:b4c05>, 0, b'd')
[352] derecho.Node<Node:b4c03>:OUTPUT: in deliver_upcall(), gi:  5  and req:  (<Sim:b4c05>, 1, b'\xe1')
[352] derecho.Node<Node:b4c04>:OUTPUT: response sent to the client/sim process, index:  2 response:  0
[352] derecho.Node<Node:b4c03>:OUTPUT: response sent to the client/sim process, index:  5 response:  1
[352] derecho.Node<Node:b4c04>:OUTPUT: in deliver_upcall(), gi:  3  and req:  None
[352] derecho.Node<Node:b4c04>:OUTPUT: in deliver_upcall(), gi:  4  and req:  (<Sim:b4c05>, 2, b'+')
[352] derecho.Sim<Sim:b4c05>:OUTPUT: ---- Received tid:  1  from:  <Node:b4c03>  with num_resp:  3
[352] derecho.Node<Node:b4c04>:OUTPUT: response sent to the client/sim process, index:  4 response:  2
[352] derecho.Node<Node:b4c04>:OUTPUT: in deliver_upcall(), gi:  5  and req:  (<Sim:b4c05>, 1, b'\xe1')
[353] derecho.Node<Node:b4c04>:OUTPUT: response sent to the client/sim process, index:  5 response:  1
[354] derecho.Node<Node:b4c04>:OUTPUT: sending no-op, case 2
[355] derecho.Node<Node:b4c03>:OUTPUT: in deliver_upcall(), gi:  6  and req:  (<Sim:b4c05>, 3, b'\x14')
[356] derecho.Node<Node:b4c03>:OUTPUT: response sent to the client/sim process, index:  6 response:  3
[356] derecho.Sim<Sim:b4c05>:OUTPUT: ---- Received tid:  3  from:  <Node:b4c03>  with num_resp:  4
[357] derecho.Node<Node:b4c04>:OUTPUT: in deliver_upcall(), gi:  6  and req:  (<Sim:b4c05>, 3, b'\x14')
[357] derecho.Node<Node:b4c04>:OUTPUT: response sent to the client/sim process, index:  6 response:  3
[367] derecho.Node<Node:b4c02>:OUTPUT: in deliver_upcall(), gi:  0  and req:  None
[367] derecho.Node<Node:b4c02>:OUTPUT: in deliver_upcall(), gi:  1  and req:  None
[367] derecho.Node<Node:b4c02>:OUTPUT: in deliver_upcall(), gi:  2  and req:  (<Sim:b4c05>, 0, b'd')
[367] derecho.Node<Node:b4c02>:OUTPUT: response sent to the client/sim process, index:  2 response:  0
[367] derecho.Node<Node:b4c02>:OUTPUT: in deliver_upcall(), gi:  3  and req:  None
[367] derecho.Node<Node:b4c02>:OUTPUT: in deliver_upcall(), gi:  4  and req:  (<Sim:b4c05>, 2, b'+')
[368] derecho.Node<Node:b4c02>:OUTPUT: response sent to the client/sim process, index:  4 response:  2
[368] derecho.Node<Node:b4c02>:OUTPUT: in deliver_upcall(), gi:  5  and req:  (<Sim:b4c05>, 1, b'\xe1')
[368] derecho.Node<Node:b4c02>:OUTPUT: response sent to the client/sim process, index:  5 response:  1
[368] derecho.Node<Node:b4c02>:OUTPUT: in deliver_upcall(), gi:  6  and req:  (<Sim:b4c05>, 3, b'\x14')
[368] derecho.Node<Node:b4c02>:OUTPUT: response sent to the client/sim process, index:  6 response:  3
[368] derecho.Node<Node:b4c02>:OUTPUT: in deliver_upcall(), gi:  7  and req:  (<Sim:b4c05>, 5, b'\x83')
[368] derecho.Node<Node:b4c02>:OUTPUT: response sent to the client/sim process, index:  7 response:  5
[368] derecho.Node<Node:b4c02>:OUTPUT: in deliver_upcall(), gi:  8  and req:  None
[368] derecho.Node<Node:b4c02>:OUTPUT: in deliver_upcall(), gi:  9  and req:  (<Sim:b4c05>, 4, b'\xbf')
[369] derecho.Sim<Sim:b4c05>:OUTPUT: ---- Received tid:  5  from:  <Node:b4c02>  with num_resp:  5
[369] derecho.Node<Node:b4c02>:OUTPUT: response sent to the client/sim process, index:  9 response:  4
[369] derecho.Node<Node:b4c02>:OUTPUT: in deliver_upcall(), gi:  10  and req:  (<Sim:b4c05>, 6, b'\xca')
[369] derecho.Node<Node:b4c02>:OUTPUT: response sent to the client/sim process, index:  10 response:  6
[369] derecho.Sim<Sim:b4c05>:OUTPUT: ---- Received tid:  4  from:  <Node:b4c02>  with num_resp:  6
[369] derecho.Node<Node:b4c02>:OUTPUT: in deliver_upcall(), gi:  11  and req:  None
[369] derecho.Node<Node:b4c02>:OUTPUT: in deliver_upcall(), gi:  12  and req:  (<Sim:b4c05>, 7, b'\x06')
[369] derecho.Node<Node:b4c02>:OUTPUT: response sent to the client/sim process, index:  12 response:  7
[369] derecho.Sim<Sim:b4c05>:OUTPUT: ---- Received tid:  6  from:  <Node:b4c02>  with num_resp:  7
[370] derecho.Sim<Sim:b4c05>:OUTPUT: ---- Received tid:  7  from:  <Node:b4c02>  with num_resp:  8
[371] derecho.Node<Node:b4c02>:OUTPUT: in deliver_upcall(), gi:  13  and req:  (<Sim:b4c05>, 9, b'0')
[371] derecho.Node<Node:b4c02>:OUTPUT: response sent to the client/sim process, index:  13 response:  9
[371] derecho.Node<Node:b4c02>:OUTPUT: in deliver_upcall(), gi:  14  and req:  None
[371] derecho.Node<Node:b4c02>:OUTPUT: in deliver_upcall(), gi:  15  and req:  (<Sim:b4c05>, 8, b'\xb3')
[371] derecho.Sim<Sim:b4c05>:OUTPUT: ---- Received tid:  9  from:  <Node:b4c02>  with num_resp:  9
[371] derecho.Node<Node:b4c02>:OUTPUT: response sent to the client/sim process, index:  15 response:  8
[372] derecho.Sim<Sim:b4c05>:OUTPUT: ---- Received tid:  8  from:  <Node:b4c02>  with num_resp:  10
[372] derecho.Sim<Sim:b4c05>:OUTPUT: ---- ---- all responded
[374] derecho.Node_<Node_:4ec01>:OUTPUT: AVG cpu_times 0.0058
[374] derecho.Node_<Node_:4ec01>:OUTPUT: AVG run_times 0.05636
[374] derecho.Node_<Node_:4ec01>:OUTPUT: AVG throughput 55.0
[374] da.api<MainProcess>:INFO: Main process terminated.
--------- time: 0.18188166618347168
\end{lstlisting}
\end{document}